\documentclass[aps,prc,twocolumn, superscriptaddress,showkeys,nofootinbib]{revtex4-1}  

\usepackage[utf8]{inputenc}
\usepackage{graphicx}
\usepackage{subcaption}

\usepackage{graphicx,color}
\usepackage{amsmath,amssymb,amsfonts}
\usepackage{hyperref}

\usepackage{xspace}	
\input{defs.tex}

\begin{document}
\title{
Studying Two-Photon Exchange in Deep Inelastic Scattering with the HERA Data
}
\medskip
\author{Henry~T.~Klest} 
\email{hklest@anl.gov}
\affiliation{Physics Division, Argonne National Laboratory, \\
Lemont, IL 60439, USA}
\begin{abstract}
Two-photon exchange (TPE) is one of the leading explanations for discrepancies in measurements of the proton electromagnetic form factors. It has been proposed that TPE could impact not only elastic scattering but also the cross sections for both inclusive deep inelastic scattering (DIS) and semi-inclusive DIS, thereby affecting the interpretation of DIS structure functions in terms of parton distributions. It is expected that higher-order QED effects such as TPE should manifest as a deviation from unity in the ratio of \epp and \emp DIS cross sections. We use the existing inclusive $e^{\pm}p$ DIS data from HERA and SLAC to constrain higher-order QED effects on inclusive DIS.
\end{abstract}
\keywords{}
\maketitle

\section{Introduction} \label{sec:intro}
Electron scattering from hadronic targets provides the cleanest process for elucidating the electromagnetic structure of the target. At Born-level, this process proceeds via the exchange of a single virtual photon. However, there presently exists a long-standing discrepancy between the Rosenbluth and polarization transfer techniques for measuring the electromagnetic form factors of the proton~\cite{JeffersonLabHallA:1999epl,Arrington:2011dn,Qattan:2024pco}. A plausible resolution to this discrepancy is two-photon exchange (TPE), since Rosenbluth measurements are generally more sensitive to TPE effects than polarization transfer~\cite{Arrington:2011dn}. TPE contributions enter the cross section with opposite sign depending on the beam charge~\cite{Arrington:2011dn}, meaning TPE can be measured by comparing the cross sections for \emp and \epp scattering. The observable we will focus on is the ratio of the positron and electron cross sections, 
\begin{align}
R_{\pm}=\frac{\sigma_{e^+}}{\sigma_{e^-}}.
\end{align}
In elastic $ep$ scattering\footnote{In elastic scattering, the quantity $R_{\pm}$ is often written as $R_{2\gamma}$.}, $R_{\pm}\approx1-2\delta_{\mathrm{TPE}}$, providing access to the contribution of TPE to the cross section, $\delta_{\mathrm{TPE}}$. The explicit definition of $\delta_{\mathrm{TPE}}$ can be found in e.g. Eq. 36 of Ref.~\cite{Arrington:2011dn}, where it is referred to as $\bar{\delta}$. In the case of DIS, $R_{\pm}$ is sensitive to higher-order QED effects, including both TPE and the interference between radiation from the lepton and from the quarks. 

Various groups have analyzed the TPE contribution to elastic scattering theoretically and phenomenologically~\cite{Fishbane:1973foj,Bodwin:1975ty,Gorchtein:2006mq,Borisyuk:2008es,Qattan:2011ke,Alberico:2009yp,Blunden:2017nby,Borisyuk:2008db,Chen:2004tw,A1:2013fsc,Tomalak:2017shs,Qattan:2024pco} to parameterize the effect of TPE on the cross section. Many of the existing parameterizations predict that the contribution of TPE grows with the four-momentum transfer to the electron\footnote{In the case of two-photon exchange in DIS, \qsq no longer represents the virtuality of a single exchanged boson. Therefore, we stick with the standard ``experimentalist's'' definition, which is simply that \qsq is the four-momentum transfer to the scattered electron.}, \qsq. A scaling of TPE effects with $\qsq$ is natural to resolve the form factor discrepancies. In elastic scattering, $R_{\pm}$ has been measured primarily at $Q^2<5~\GeVsq$~\cite{OLYMPUS:2016gso,Rachek:2014fam,CLAS:2016fvy,Tomasi-Gustafsson:2009cfi,Camilleri:1969ah,Fancher:1976ea,Mar:1968qd,Bouquet:1968yqa}. The results suggest a mild increase in the value of $R_{\pm}$ at larger \qsq where partonic processes are expected to contribute~\cite{Borisyuk:2008db,Afanasev:2005mp}. Therefore, the issue of TPE may not be limited only to the elastic case~\cite{Afanasev:2012jq,Afanasev:2023gev,Cao:2019vkw,Lee:2025dsz}. 

\begin{figure}[h]
    \centering
    \qquad 
    \includegraphics[width=0.7\linewidth,
        trim={0.1cm 0.1cm 0.1cm 0.1cm},
        clip
    ]{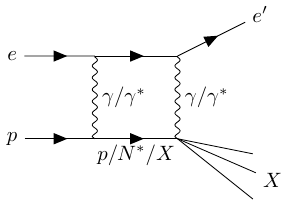}
    \caption{Example TPE diagram contributing to inclusive DIS.}
    \label{fig:Feynman}
\end{figure}

There is no obvious reason why TPE would be large in elastic scattering and small in inelastic scattering, and recent calculations have shown potentially significant effects for exclusive pion electroproduction~\cite{Afanasev:2012jq,Cao:2019vkw} and semi-inclusive DIS~\cite{Lee:2025dsz}. DIS data provide the backbone for extractions of proton parton distribution functions (PDFs). If higher-order QED effects such as the TPE diagram shown in Fig.~\ref{fig:Feynman} play an important role in the DIS cross section, it could be that existing extractions of parton distributions erroneously attribute QED effects to the parton distributions themselves. With precise data on $e^-p$ inclusive DIS cross sections across a broad range of \xbj, \qsq, and $\varepsilon$ expected from the Electron-Ion Collider (EIC)\footnote{The EIC project presently does not include a positron beam option.}, understanding the size and scaling of higher-order QED effects in DIS is vital for a robust interpretation of the measured DIS structure functions in terms of parton distributions. To explain the form factor discrepancy, TPE must scale with the photon polarization parameter
\begin{align}
\varepsilon=\frac{1-y-\frac{Q^2}{4E^2}}{1-y+\frac{y^2}{2}+\frac{Q^2}{4E^2}},
\end{align} 
where $E$ is the electron beam energy and $y$ is the fraction of $E$ carried by the scattered electron, both defined in the target rest frame. Since $\varepsilon$ varies rapidly at high $y$ and thus low-\xbj, TPE in DIS could easily obfuscate low-\xbj measurements targeting gluon saturation, one of the pillars of the EIC science program~\cite{AbdulKhalek:2021gbh}. 

Data on higher-order QED effects in DIS on nucleon targets are limited. The HERMES collaboration published an explicit search for TPE in DIS on a polarized proton target in Ref.~\cite{HERMES:2009hsi}. In that result, the target transverse single-spin asymmetries, which vanish at Born-level but are non-zero in the presence of TPE~\cite{Metz:2012ui,Goity:2023sph}, are measured to be consistent with zero at the level of 0.1\% for $1<\qsq<20~\GeVsq$. A result from Hall A at Jefferson Lab~\cite{Katich:2013atq} measures the same target transverse spin asymmetry in DIS on polarized $^3\mathrm{He}$ and observes a 2.89$\sigma$ deviation from zero for $1<\qsq<4~\GeVsq$, which can be interpreted as evidence for TPE in DIS on the neutron of a magnitude similar to that predicted by Ref.~\cite{Metz:2012ui}. The only results on the beam charge asymmetry $R_{\pm}$ in DIS on nucleon targets are from SLAC~\cite{Rochester:1975fk,Fancher:1976ea} in the range $1.2<\qsq<15~\GeVsq$. No clear non-zero value of $R_{\pm}$ is observed within the uncertainties in either measurement. 

The goal of this paper is to perform a simple phenomenological study of $R_{\pm}$ in the combined inclusive DIS data from the HERA collider experiments. Since H1 and ZEUS both collected roughly equal statistics for \emp and \epp scattering at values of \qsq up to 50,000 \GeVsq, the HERA data provide an ideal testing ground for the \qsq and $\varepsilon$ scaling of higher-order QED effects.

\section{Dataset and Kinematic Selection}
\label{sec:analysis}
To investigate higher-order QED effects in DIS, we utilize the dataset of Ref.~\cite{H1:2015ubc,hepdata.68951}, namely the final combined inclusive DIS cross sections from H1 and ZEUS at $\sqrt{s}=319$ GeV. Over the lifetime of HERA from 1992 to 2007, both H1 and ZEUS each collected about 500 pb$^{-1}$ of integrated luminosity split roughly equally between \emp and \epp. The combined data benefit from the fact that H1 and ZEUS measured the cross sections with independent detectors and different analysis techniques, thereby significantly improving the precision on the combined results. The typical total uncertainty on the cross sections is $1-5\%$. The large number of precise data points across a wide range of \qsq and $\varepsilon$ make this dataset well-suited for studying higher-order QED effects. We restrict our analysis to the region where the data are presented with identical binning in \qsq and \xbj for $e^+p$ and $e^-p$, namely $\qsq>60~\GeVsq$. Unfortunately, the \emp cross section was not measured at lower values of \qsq by either H1 or ZEUS, in spite of the large available dataset\footnote{The H1 and ZEUS datasets are archived according to the Data Preservation in HEP framework~\cite{South:2012vh,DPHEPStudyGroup:2012dsv}. As a result, a re-analysis of the lower $\qsq$ inclusive DIS beam charge asymmetry remains possible.}. Our goal in this section is to define the optimal subset of the HERA data points to isolate potential TPE effects. The uncertainties on the $e^-p$ and $e^+p$ cross section data points are provided as statistical, uncorrelated, and correlated uncertainties. We simply combine these uncertainties in quadrature, in lieu of a publicly available covariance matrix between the $e^-p$ and $e^+p$ datasets.

\begin{figure}[h!]
    \centering
    \qquad 
    \includegraphics[width=1\linewidth]{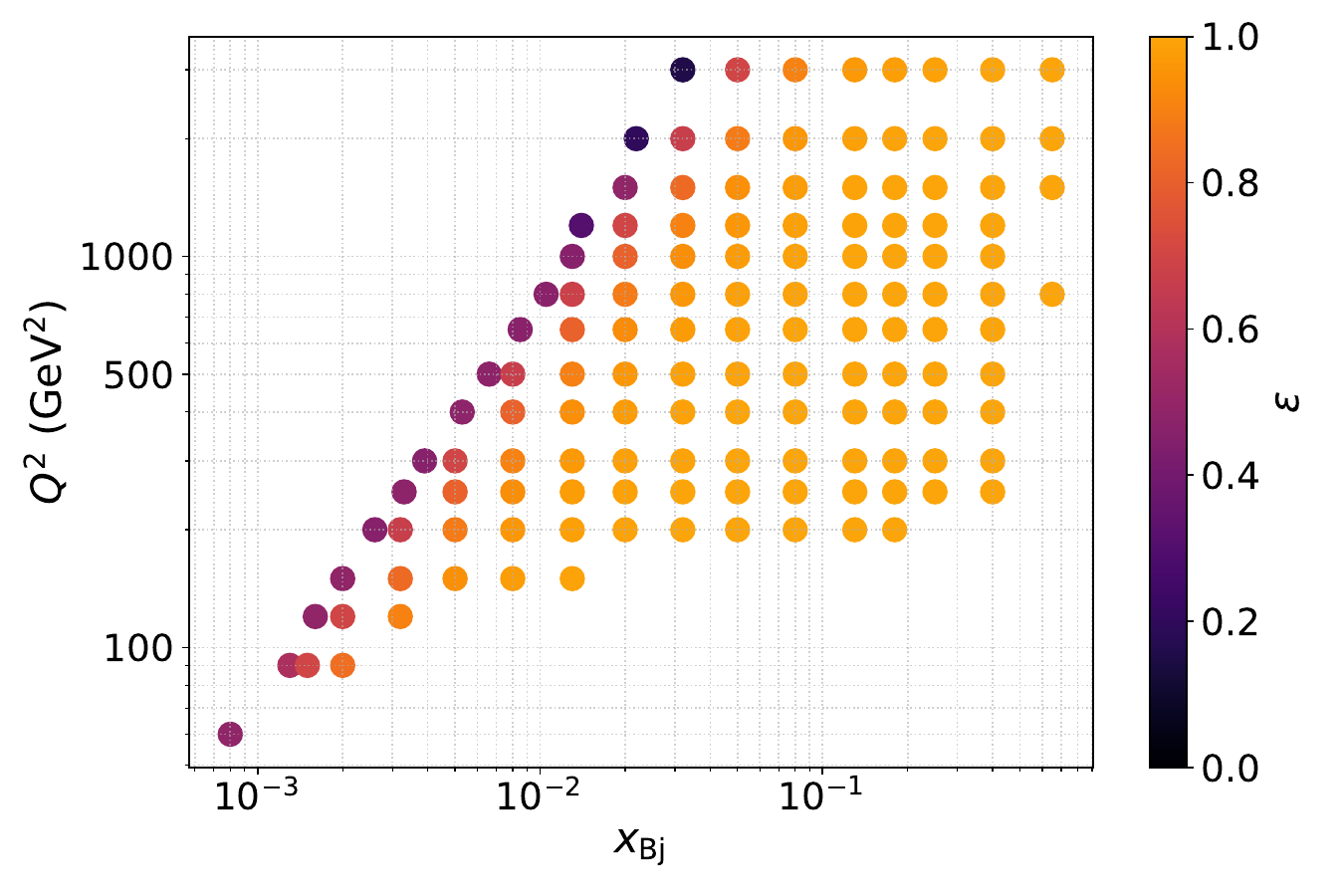}
    \caption{HERA neutral current DIS data points in the kinematic plane of \xbj and \qsq. This subset of the data corresponds to the bins where the $e^+p$ and $e^-p$ cross sections are presented with the same binning. Only data points with $\qsq\leq3000~\GeVsq$ are shown. The color scale shows the value of $\varepsilon$ at the kinematic point.}
    \label{fig:xQ2}
\end{figure}

Electroweak effects produce substantial deviations from unity for the ratio $\frac{\sigma_{e^+}}{\sigma_{e^-}}$ at large \qsq. This feature has been used to extract the $F_3^{\gamma Z}$ structure function~\cite{H1:2015ubc}, which provides access to the behavior of the valence quark distributions at lower-\xbj. The neutral current cross section for \epp falls faster than \emp at $Q^2>1000$ \GeVsq due to $\gamma Z$ interference, producing values of $\frac{\sigma_{e^+}}{\sigma_{e^-}}$ which are less than one and have a non-linear dependence on $\varepsilon$. $\gamma Z$ interference also means that $\frac{\sigma_{e^+}}{\sigma_{e^-}}$ need not equal one at $\varepsilon=1$. In contrast, TPE is expected to produce a value of $\frac{\sigma_{e^+}}{\sigma_{e^-}}$ which is equal to one at $\varepsilon=1$. 

\begin{figure}[h!]
  \centering
  \includegraphics[width=1\linewidth,
      trim={0.8cm 1cm 2cm 4cm},clip
  ]{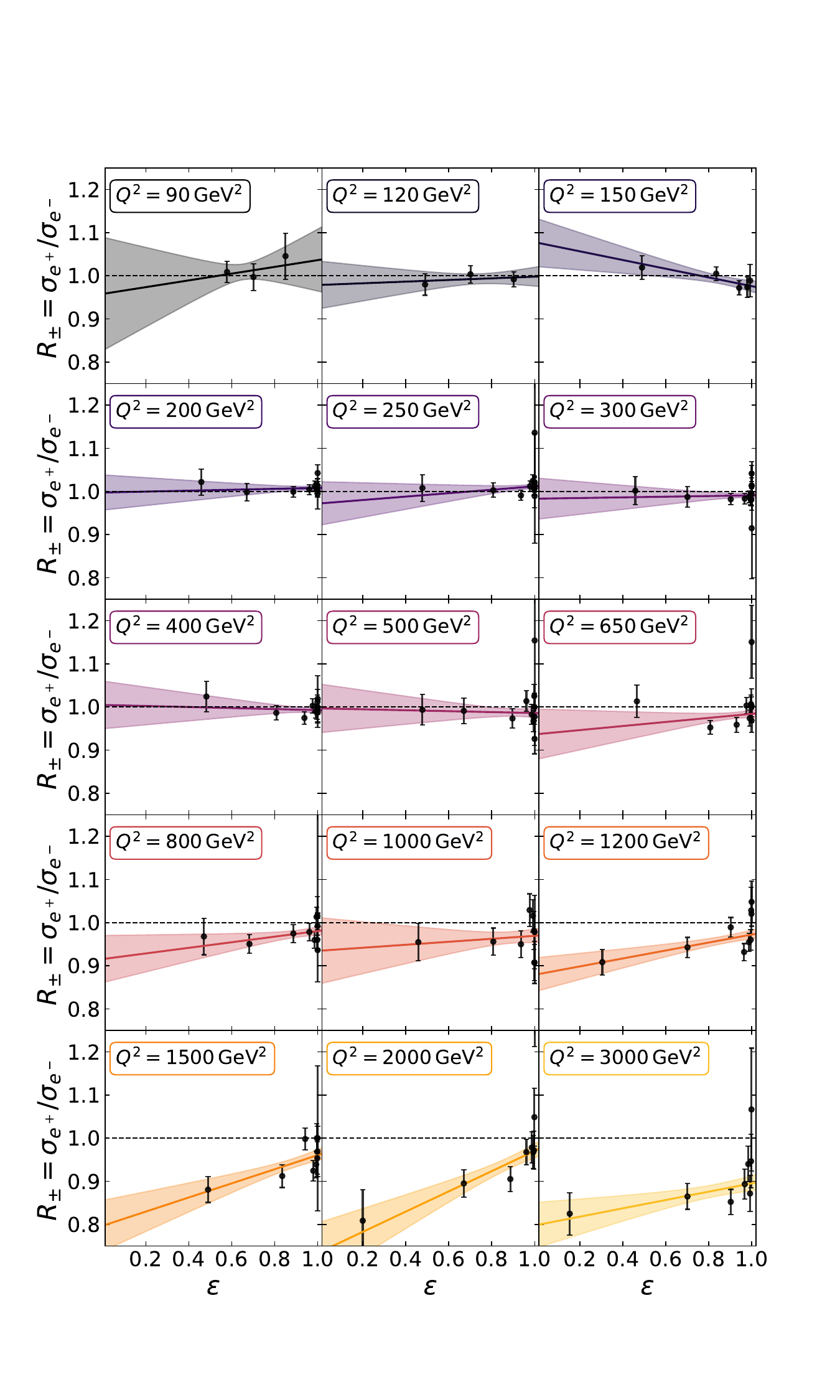}

  \vspace{0em} 

  \includegraphics[width=1\linewidth]{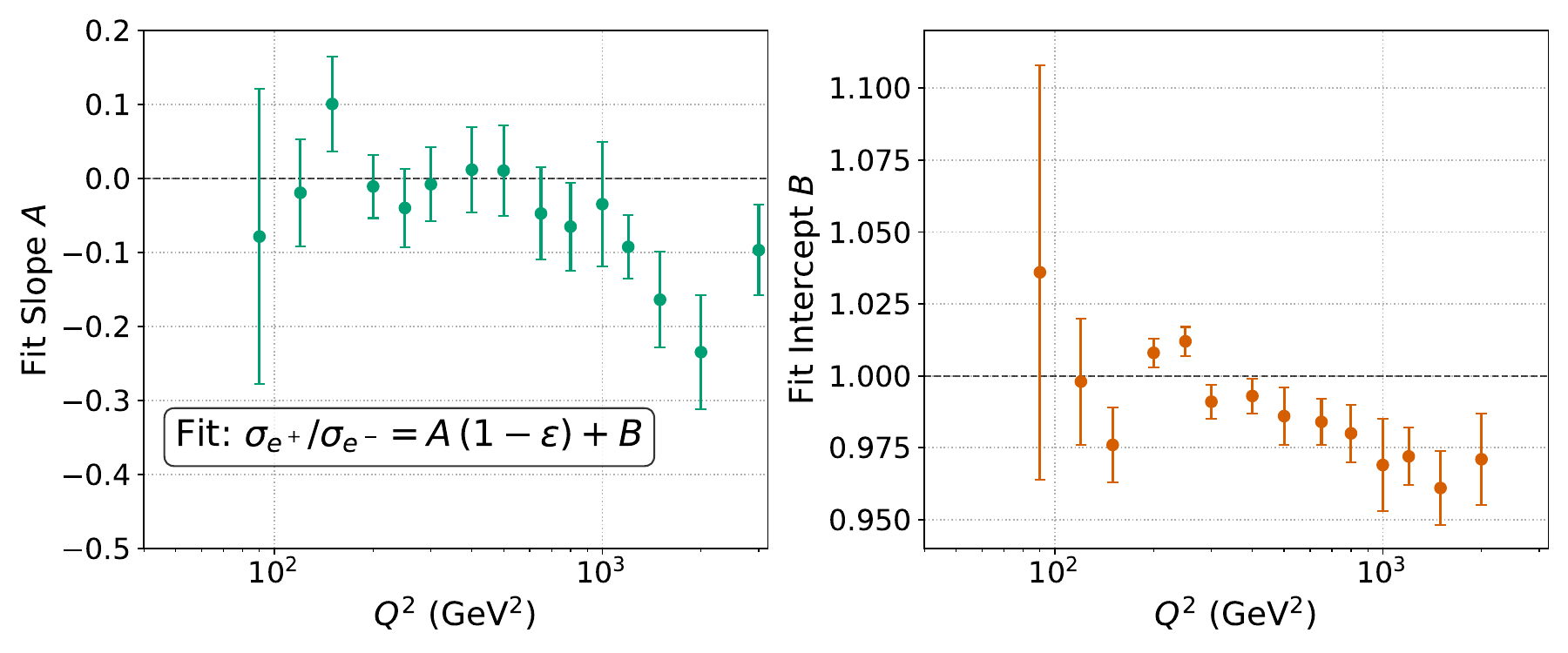}

  \caption{\textbf{Top:} Linear fits $R_{\pm}(\varepsilon)=A(1-\varepsilon)+B$ to the data in different bins of $Q^2$. \textbf{Bottom:} Extracted fit parameters. The existence of a negative slope, $A$, and an intercept, $B$, less than one indicate the effect of electroweak interference.}
  \label{fig:combined_vertical}
\end{figure}

To select the region of \qsq to use for studying higher-order QED effects, the data were analyzed for signs of electroweak effects. Since the elastic form factor discrepancies exist at lower \qsq, we expect that any higher-order QED effect should not turn on at the large values of \qsq available in the HERA dataset. Therefore, we can ascribe any appearance of a trend in the data at higher \qsq to electroweak effects. To define a reasonable range of \qsq free of electroweak contamination, the $\varepsilon$ distributions of all the \qsq bins were fit with the function $A(1-\varepsilon)+B$. The results of the fit are presented in Fig.~\ref{fig:combined_vertical}. It can be seen from Fig.~\ref{fig:combined_vertical} that the slope $A$ of the $\varepsilon$ distributions becomes non-negative in the region of $\qsq\approx500~\GeVsq$, while the intercept $B$ approximately reaches unity for $\qsq\approx300~\GeVsq$. We therefore choose a conservative range and restrict our analysis to $Q^2\leq300$ \GeVsq to avoid the impact of electroweak effects on $R_{\pm}$. 

How the data were corrected for radiative effects can obviously affect their interpretation in terms of two-photon exchange. The nominal task of a radiative correction is to correct the data to the Born-level, where extra QED effects are not present. Therefore, it is prudent to ask whether the flatness of $R_{\pm}$ is simply related to the fact that higher-order QED effects were corrected out by the radiative correction procedure applied to the data. Both H1 and ZEUS used the HERACLES program~\cite{Kwiatkowski:1990es,Charchula:1994kf} to perform radiative corrections. HERACLES includes $\mathcal{O}(\alpha_{EM})$ QED corrections for photon radiation from the lepton, as well as one-loop virtual corrections. At the time of the HERA NC DIS analyses, the contributions of two-photon exchange and radiation from the quarks had been calculated~\cite{Bardin:1988by,Arbuzov:1995id}. However, two-photon exchange was not corrected for in the inclusive DIS analyses~\cite{H1:2018mkk,H1:2000olm,ZEUS:2016vyd}, which made use of the DJANGOH~\cite{Charchula:1994kf} interface between HERACLES and LEPTO with two-boson exchange turned off. Therefore the data of Ref.~\cite{H1:2015ubc} can indeed be used to infer the size of these higher-order QED effects.

\section{Results and Discussion}
\label{sec:results}
\begin{figure*}[h!]
  \centering
  \begin{subfigure}[b]{0.48\linewidth}
    \includegraphics[width=\linewidth]{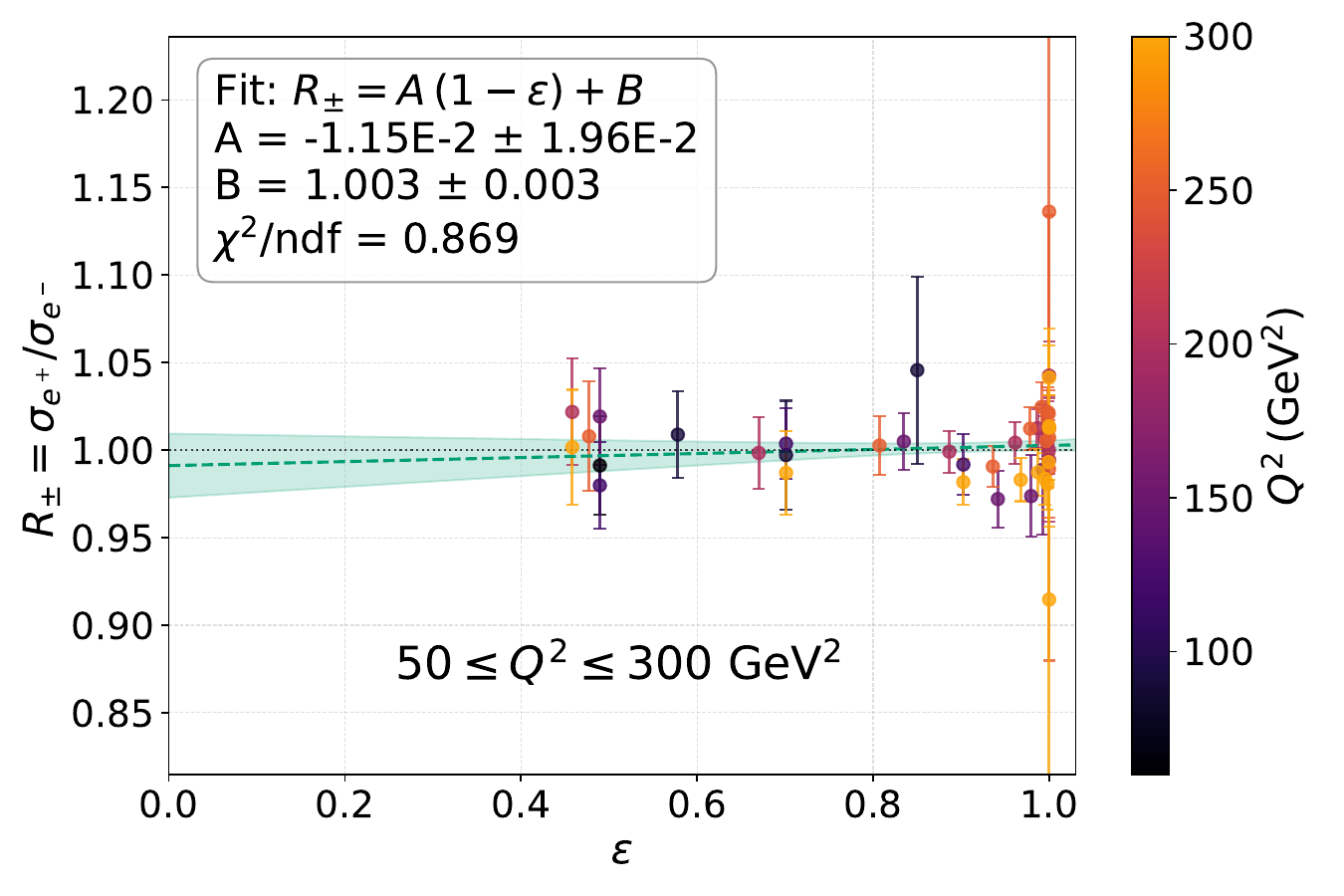}
    \label{fig:R_v_eps}
  \end{subfigure}
  \hfill
  \begin{subfigure}[b]{0.48\linewidth}
    \includegraphics[width=\linewidth]{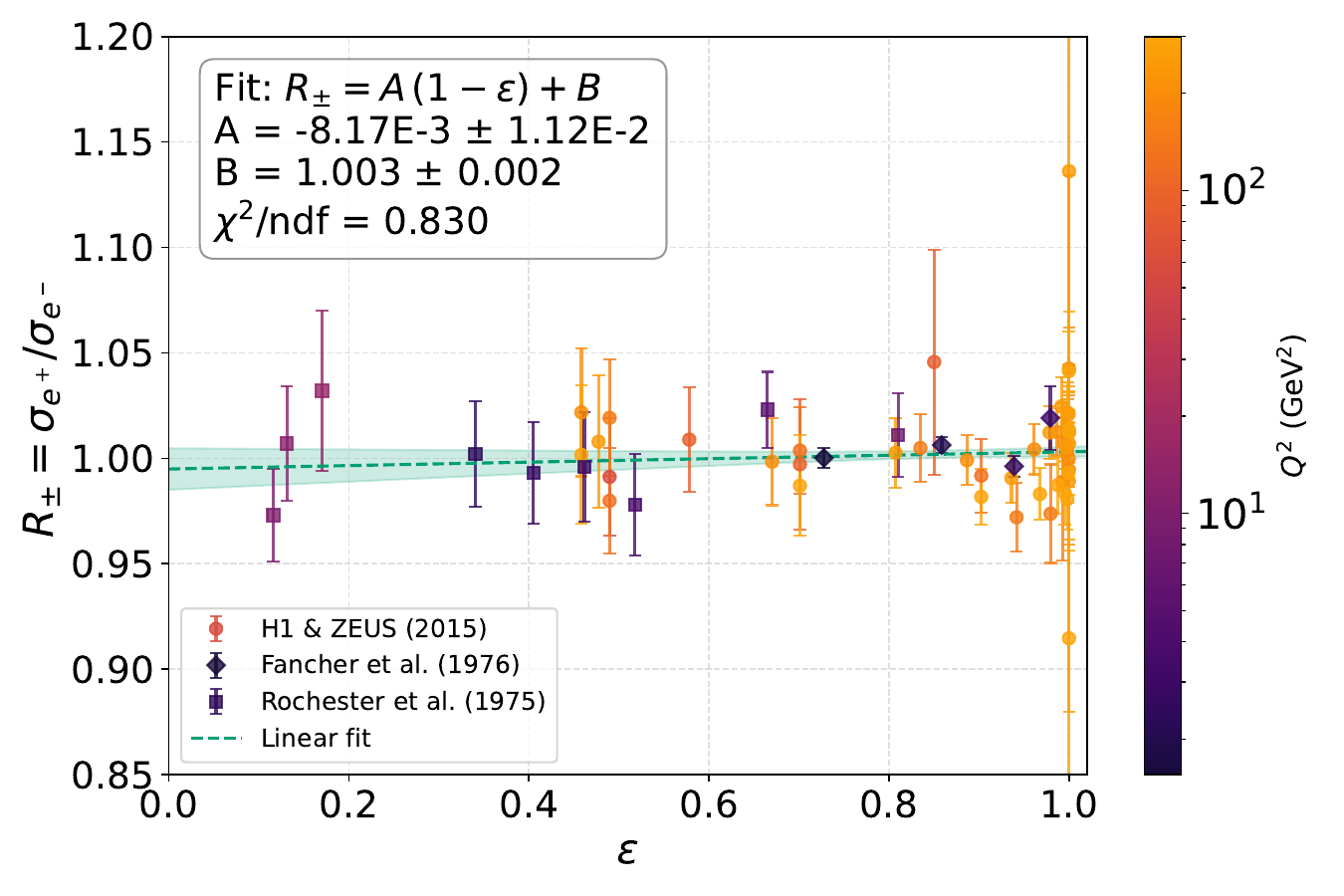}
    \label{fig:R_v_eps_GF}
  \end{subfigure}
  \caption{Linear fits of $R_{\pm}$ as a function of $\varepsilon$.  \textbf{Left:} Fit to the HERA data only, in the range $60\leq\qsq\leq300\ \GeVsq$.  \textbf{Right}: Global fit including the SLAC results from Refs.~\cite{Rochester:1975fk,Fancher:1976ea}. Note the different color scales between panels.}
  \label{fig:combined_epsilon}
\end{figure*}

\begin{figure*}[h!]
  \centering
  \begin{subfigure}[b]{0.48\linewidth}
    \includegraphics[width=\linewidth]{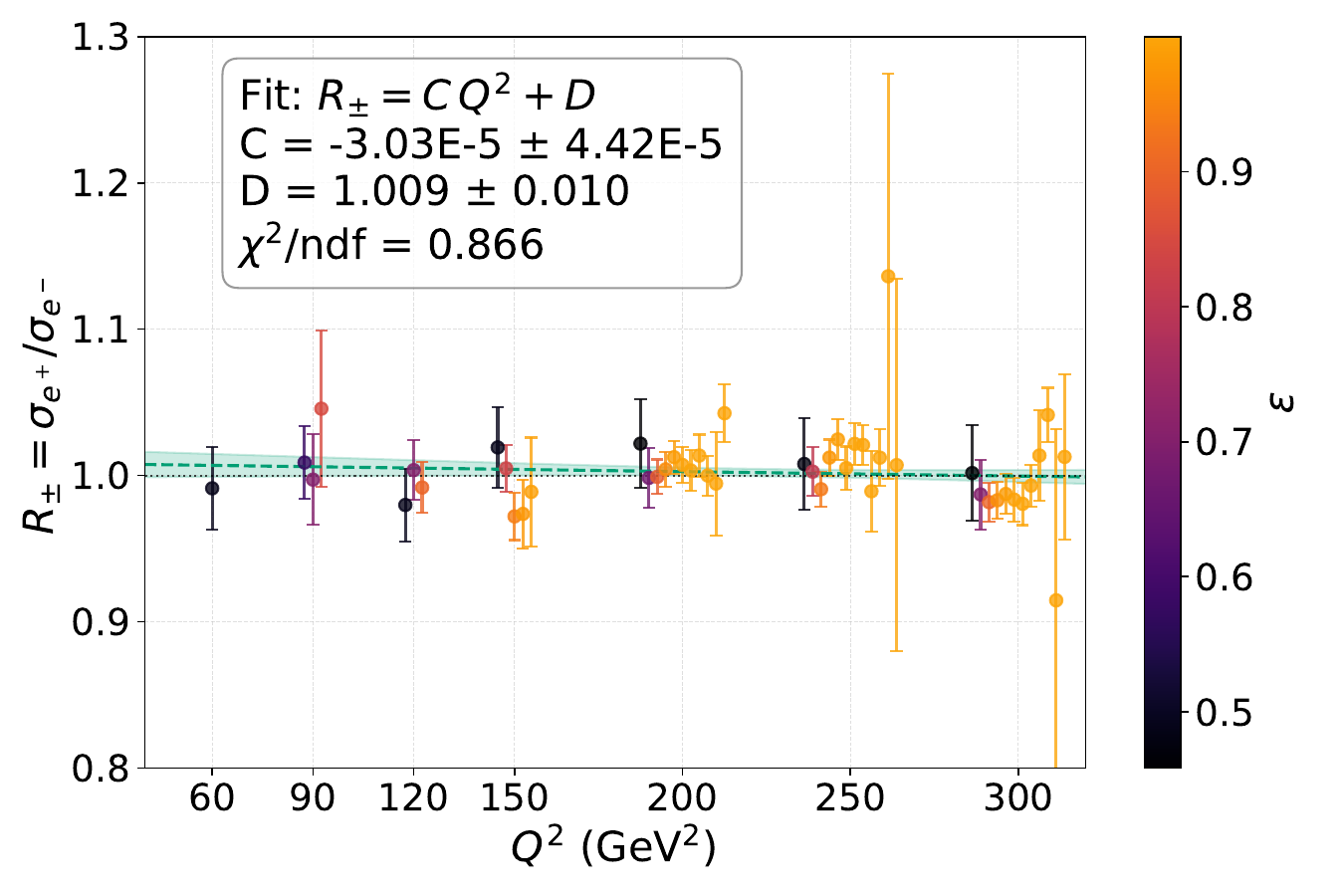}
    \label{fig:R_v_Q2}
  \end{subfigure}%
  \hfill
  \begin{subfigure}[b]{0.48\linewidth}
    \includegraphics[width=\linewidth]{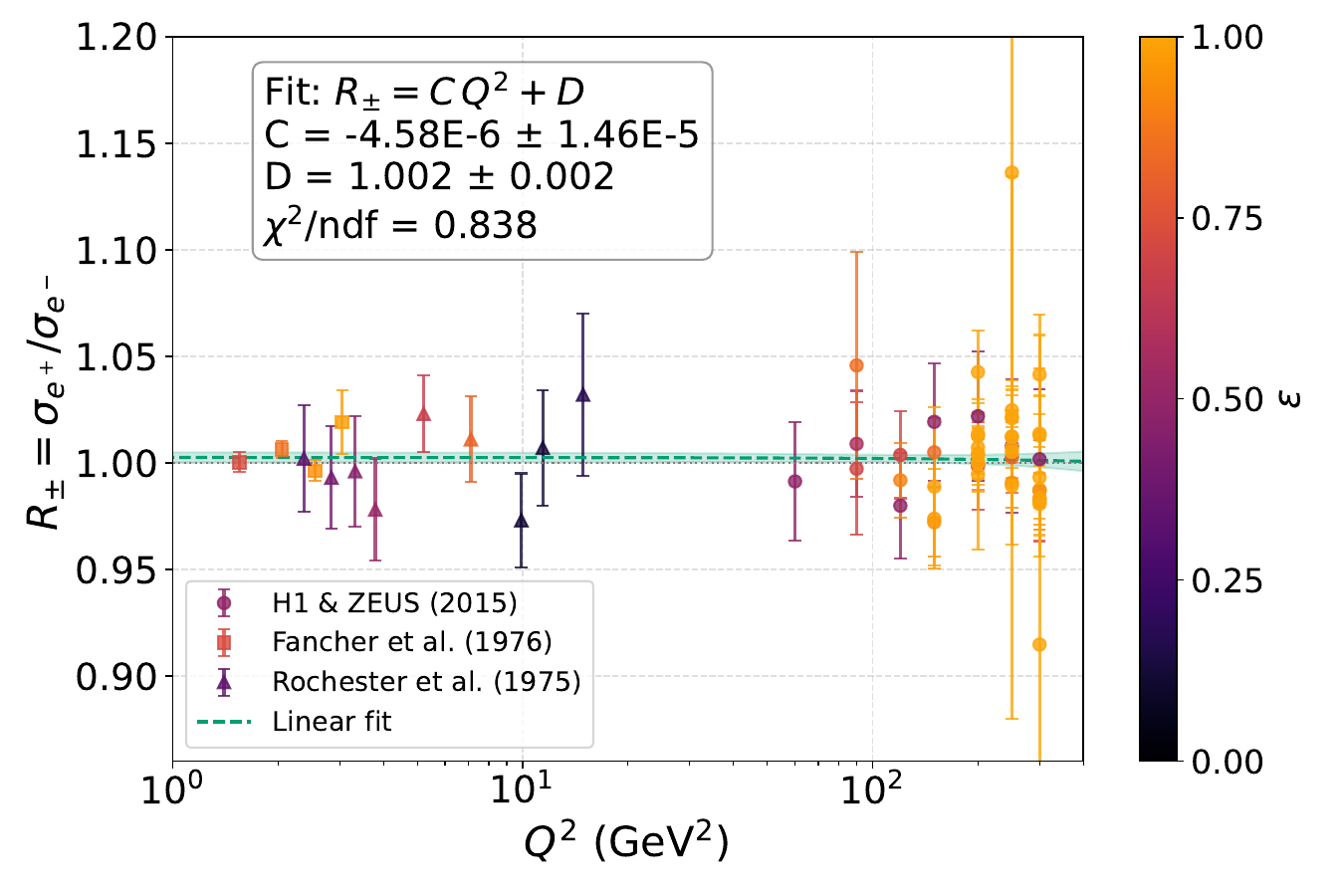}
    \label{fig:R_v_Q2_GF}
  \end{subfigure}
  \caption{Linear fits of $R_{\pm}$ as a function of $Q^2$.  
    \textbf{Left:} HERA data only, points at the same \qsq are offset horizontally for clarity. \textbf{Right:} Global fit including the SLAC~\cite{Fancher:1976ea,Rochester:1975fk} measurements.}
  \label{fig:combined_Q2}
\end{figure*}

In general, the values of $R_{\pm}$ for the data with $\qsq\leq300~\GeVsq$ are in good agreement with unity. In lieu of any obvious trend, we perform linear fits to $R_{\pm}(\varepsilon)$ and $R_{\pm}(\qsq)$. Fits to the single-dimensional $\varepsilon$  and \qsq distributions for the $\qsq\leq300~\GeVsq$ data are shown in Figs.~\ref{fig:combined_epsilon} and ~\ref{fig:combined_Q2}, respectively. No significant slopes are observed in $R_{\pm}$ as functions of $\varepsilon$ or \qsq, setting limits on the dependencies and magnitude of higher-order QED effects in DIS. A linear fit to the \xbj distribution similarly revealed no slope.

The weighted average value of the HERA $R_{\pm}$ data in the kinematic region of $60\leq\qsq\leq300~\GeVsq$ is $1.0018\pm0.0025$. When the data of Refs.~\cite{Fancher:1976ea,Rochester:1975fk} are included, the weighted average value of $R_{\pm}$ data in the region $1.2\leq\qsq\leq300~\GeVsq$ becomes $1.0019 \pm 0.0018$. Integrating over \qsq and $\varepsilon$, these data together exclude a global value of $R_{\pm}=1.01$ at the level of $4.5\sigma$ and $R_{\pm}=0.99$ at the level of $6.6\sigma$. 

In Figs.~\ref{fig:combined_Q2} and~\ref{fig:combined_epsilon} we perform a global fit to the existing data on $R_{\pm}(\qsq)$ and $R_{\pm}(\varepsilon)$in DIS on proton targets. The SLAC data on $R_{\pm}$ significantly improve the lever arms for the fits. The data taken together show a clear lack of a dependence on \qsq and $\varepsilon$. An $R_{\pm}(\qsq)$ that linearly increases by 0.01\% per \GeVsq of \qsq is ruled out at the level of $6.6\sigma$.

\begin{table}[htbp]
  \centering
    \scriptsize 
  \begin{tabular}{llcc}
    \hline
    \textbf{Fit form} 
      & \textbf{Data set} 
      & \textbf{Slope} 
      & \textbf{Intercept} \\
    \hline
    $A\,(1-\varepsilon) + B$ 
      & HERA 
      & $-1.15\cdot10^{-2}\pm1.96\cdot10^{-2}$ 
      & $1.003\pm0.003$ \\
    $A\,(1-\varepsilon) + B$ 
      & Global 
      & $-8.17\cdot10^{-3}\pm1.12\cdot10^{-2}$ 
      & $1.003\pm0.002$ \\
    $C\,Q^2 + D$ 
      & HERA
      & $-3.03\cdot10^{-5}\pm4.42\cdot10^{-5}$ 
      & $1.009\pm0.010$ \\
    $C\,Q^2 + D$ 
      & Global 
      & $-4.58\cdot10^{-6}\pm1.46\cdot10^{-5}$ 
      & $1.002\pm0.002$ \\
    \hline
  \end{tabular}
  
    \normalsize 
    \caption{Linear fit parameters for $R_{\pm}$ vs.\ $Q^2$ and $\varepsilon$.}
  \label{tab:fit-summary}
\end{table}

\begin{figure*}[t!]
  \centering
  \begin{subfigure}[b]{0.48\linewidth}
    \includegraphics[width=\linewidth]{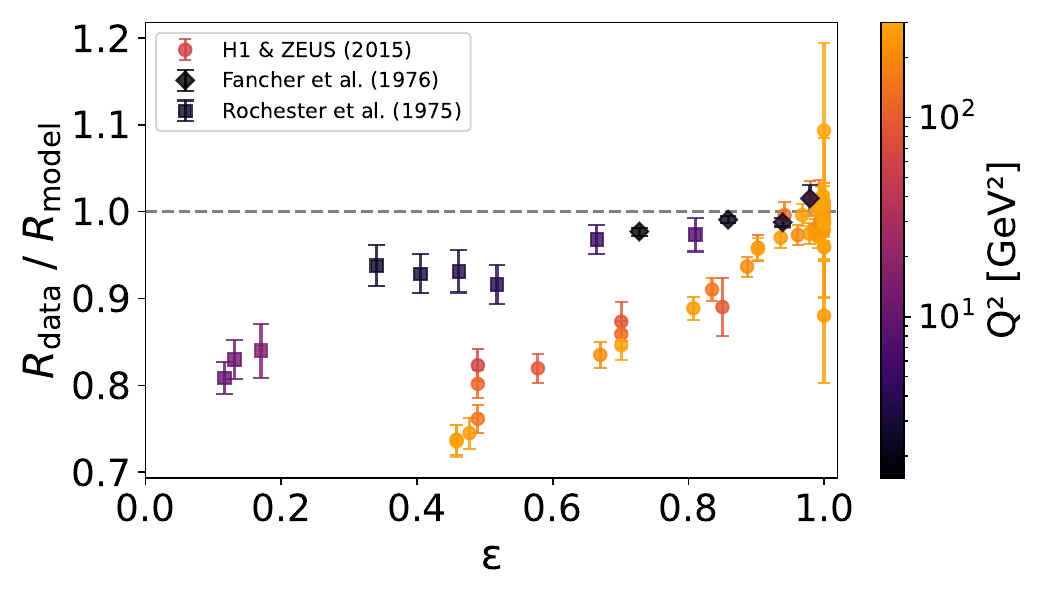}
    \label{fig:R_v_eps}
  \end{subfigure}
  \hfill
  \begin{subfigure}[b]{0.48\linewidth}
    \includegraphics[width=\linewidth]{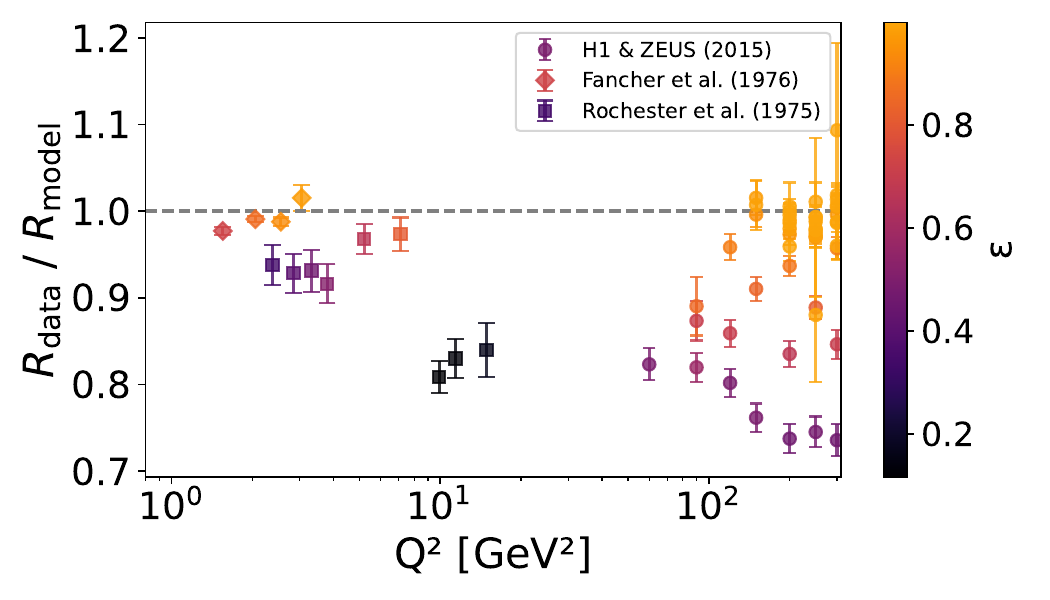}
    \label{fig:R_v_eps_GF}
  \end{subfigure}
  \caption{Comparison between data and the parameterization $R_{\pm} = 1+2\cdot 0.069\cdot(1-\varepsilon)\mathrm{ln}(0.394/\mathrm{GeV}^2\cdot Q^2+1)$. \textbf{Left:} Ratio of $R_{\pm}$ values in the data to the predicted value from the parameterization as a function of $\varepsilon$. \textbf{Right:} Same as the left panel as a function of $\qsq$.} 
  \label{fig:param}
\end{figure*}

One feature visible in Fig.~\ref{fig:combined_epsilon} is a local dip below unity in the range of $0.85\lesssim\varepsilon\lesssim0.98$. The local significance of this dip is around $2\sigma$ and the global significance is around 1$\sigma$. Interestingly, this dip does reside in a similar region of $\varepsilon$ as the $R_{\pm}<1$ region observed in the OLYMPUS elastic scattering data~\cite{OLYMPUS:2016gso}. This feature is predicted in elastic scattering by the phenomenological models of Refs.~\cite{A1:2013fsc} and~\cite{Guttmann:2010au}. The GPD-based prediction of Ref.~\cite{Afanasev:2005mp} calculates TPE for $eq$ scattering and does predict a qualitatively similar shape for the distribution of $R_{\pm}$ at large values of $\varepsilon$. However, given the limited statistical significance of the dip in the data and the fact that the model of Ref.~\cite{Afanasev:2005mp} predicts that at large \qsq a larger region of $\varepsilon$ should have $R_{\pm}<1$, we cannot draw a strong conclusion from this comparison.

The lack of a deviation from unity in $R_{\pm}$ is at first glance in tension with the asymmetry observed in Ref.~\cite{Katich:2013atq}, which can be ascribed to two-photon exchange. Since asymmetries probe the imaginary part of the TPE amplitude~\cite{Metz:2006pe} while $R_{\pm}$ probes the real part, it could be the case that there are cancellations between TPE and quark-lepton radiation interference which suppress the real part of the amplitude in the kinematic region studied by experiments thus far. An additional explanation is that the non-zero asymmetry could arise from unexpected nuclear effects in $^3$He. 


At large $Q^2$, the dominant hard two-photon exchange subprocess is expected to be the same lepton–quark box diagram in both elastic scattering and DIS, so the magnitude of TPE effects and their dependence on $Q^2$ and $\varepsilon$ could be expected to be similar across processes. To test this hypothesis, we compare the DIS data to a commonly used phenomenological parameterization of $R_{\pm}$ in elastic $ep$ scattering presented in Refs.~\cite{A1:2013fsc} and~\cite{Afanasev:2023gev}. This parameterization is obtained from a global fit to elastic $R_{\pm}$ data and thus provides a compact proxy for the $Q^2$- and $\varepsilon$-scaling of $R_{\pm}$ observed in elastic scattering. The elastic-scattering parameterization takes the form
\begin{align}
R_{\pm} = 1+2\cdot a\cdot(1-\varepsilon)\ln(b\cdot Q^2+1),
\end{align}
where $a=0.069$ and $b=0.394~\mathrm{GeV}^{-2}$.
Fig.~\ref{fig:param} shows the ratio of the existing experimental DIS data points to this parameterization\footnote{We ignore the Feshbach term since it is numerically irrelevant for the high-energy data we consider here.}. The DIS data exhibit clear tension with the elastic parameterization, which predicts values of $R_{\pm}$ as large as $\sim 1.4$ for part of the HERA kinematics. Thus beam-charge-asymmetric effects on the inclusive DIS cross section do not have similar magnitude or scaling to those in elastic scattering and the dominant mechanisms governing the value of $R_{\pm}$ do indeed appear to be significantly process-dependent.

\section{Conclusion}
\label{sec:conclusion}
In summary, the existing world data overall do not show a substantial charge asymmetry in inclusive DIS. The largest local deviation from unity is around 2$\sigma$ and is localized to the region of $0.85\lesssim\varepsilon\lesssim0.98$. The world data rule out values of $R_{\pm}$ in DIS of sizes similar to those predicted from parameterizations of elastic scattering data and set limits on the size of higher-order QED effects. QCD analyses using inclusive DIS data are likely safe from pollution by higher-order QED effects within the uncertainties of the data. 

Two possibilities arise to explain the data. One possibility is that the effects of TPE and the interference of radiation from leptons and quarks are both small in DIS. The other possibility is that in DIS there is a fortuitous cancellation between the effects of two-photon exchange and quark-lepton radiation interference that produces a value of $R_{\pm}$ that agrees very nearly with unity. Theoretical calculations for the impact of radiation from quarks on $R_{\pm}$ can help discriminate between these two possibilities.

The striking consistency with $R_{\pm}=1$ is surprising in light of the asymmetry observed in DIS on polarized $^3\mathrm{He}$ in Hall A~\cite{Katich:2013atq}. The clarification of this discrepancy and the dip around $0.85\lesssim\varepsilon\lesssim0.98$ would benefit from theoretical input as well as additional data on $R_{\pm}$ and transverse-spin asymmetries in DIS at intermediate $Q^2$ and $\epsilon$, perhaps from COMPASS/AMBER~\cite{COMPASS:2007esq,Adams:2018pwt,COMPASS:2007rjf,dHose:HP2030} or Jefferson Lab~\cite{Bernauer:2021vbn,Accardi:2020swt}.

\section*{Acknowledgements}
Thanks to A. Afanesev, J. Bernauer, J. Blümlein, S. Joosten, P. Reimer, A. Schmidt, H. Spiesberger, and M. Żurek for helpful comments and discussions. We also thank L. DeWitt for editing this manuscript.
This work was supported by the U.S. Department of Energy under Contract No. DE-AC02-06CH11357.

\clearpage
\bibliography{ref} 
\end{document}